\documentclass[preprint,showpacs,preprintnumbers,amsmath,amssymb]{revtex4}
\usepackage{amsmath,amssymb,graphics,epsfig,subfigure}
\usepackage{color}

\begin{document}
\newcommand {\nn}    {\nonumber}
\renewcommand{\baselinestretch}{1.3}

\title{Charged Spinning Black Holes as Particle Accelerators}
\author{Shao-Wen Wei\footnote{weishaow06@lzu.cn},
        Yu-Xiao Liu\footnote{liuyx@lzu.edu.cn, corresponding author},
        Heng Guo\footnote{guoh06@lzu.cn},
        Chun-E Fu\footnote{fuche08@lzu.cn}}.
 \affiliation{Institute of Theoretical Physics, Lanzhou University, Lanzhou 730000,
             People's Republic of China}

\begin{abstract}
It has recently been pointed out that the spinning Kerr black hole
with maximal spin could act as a particle collider with arbitrarily
high center-of-mass energy. In this paper, we will extend the result
to the charged spinning black hole, the Kerr-Newman black hole. The
center-of-mass energy of collision for two uncharged particles
falling freely from rest at infinity depends not only on the spin
$a$ but also on the charge $Q$ of the black hole. We find that an
unlimited center-of-mass energy can be approached with the
conditions: (1) the collision takes place at the horizon of an
extremal black hole; (2) one of the colliding particles has critical
angular momentum; (3) the spin $a$ of the extremal black hole
satisfies $\frac{1}{\sqrt{3}}\leq \frac{a}{M}\leq 1$, where $M$ is
the mass of the Kerr-Newman black hole. The third condition implies
that to obtain an arbitrarily high energy, the extremal Kerr-Newman
black hole must have a large value of spin, which is a significant
difference between the Kerr and Kerr-Newman black holes.
Furthermore, we also show that, for a near-extremal black hole, there
always exists a finite upper bound for center-of-mass energy, which decreases
with the increase of the charge $Q$.
\end{abstract}



\maketitle

\section{Introduction}

Recently, Ba\~nados, Silk and West (BSW) \cite{Banados} showed that
spinning black {holes} can play the role of particle accelerators.
Compared with terrestrial accelerators, they have a fascinating and
important property that two particles (for example, the dark matter
particles) falling freely from rest at infinity can collide with
arbitrarily high center-of-mass (CM) energy at the horizon of an
extremal Kerr black hole, which could provide a visible probe of
Planck-scale physics. However, fine-tunings arise, namely, the black
hole must be a maximally spinning one and one of the particles
should have orbital angular momentum per unit rest mass $l=2$
corresponding to marginally bound geodesics. Subsequently, in
\cite{Berti} and \cite{Jacobson}, the authors further elucidated the
mechanism for the results of BSW. They pointed out that there must
exist a practical limitation on the achievable CM energy from the
astrophysical limitations, i.e., the maximal spin, back-reaction
effects or gravitational radiation. For example, the spin $a$ of
astrophysical black holes should not exceed $\frac{a}{M}=0.998$
($M$ is the mass of an astrophysical black hole) according to
the work of Thorne \cite{Thorne}. Denoting the deviation of the spin
from its maximal value as $\epsilon=1-a$, Jacobson and Sotiriou got
the maximal CM energy \cite{Jacobson}
\begin{eqnarray}
 \frac{E_{\text{cm}}^{max}}{m_{0}}\sim 4.06 \epsilon^{-1/4}
                       +\mathcal{O}(\epsilon^{1/4}),\label{ECMPP}
\end{eqnarray}
where, $m_{0}$ is the rest mass of the colliding particles. Taking
$\frac{a}{M}=0.998$ as a limit, one will obtain the
maximal CM energy per unit rest mass 19.20, which is a finite value. Lake
also showed that the CM energy of collision at the inner horizon of
a nonextremal Kerr black hole is limited \cite{Lake}. In Ref.
\cite{Grib}, scattering of particles in gravitational field and
extraction of energy from a rotating black hole was investigated.

It is known that, the motion of a particle traveling in the
background of a charged spinning black hole depends not only on the
spin but also on the charge of the black hole. Therefore, the CM
energy of collision will also depend both on the spin and charge.
Note that there is no work focusing on the CM energy for the
collision in the background of a charged spinning black hole. So, it
is worthwhile to study the detailed behavior of the CM energy for the
collision in the background of a charged spinning black hole. For
the purpose, we will study the CM energy in the background of a
Kerr-Newman (KN) black hole. Besides the spin $a$, the black hole
has another parameter, the charge $Q$, which should affect the CM
energy. On the other hand, it is generally thought that the black
holes are surrounded by relic cold dark matter density spikes and
there exists no electromagnetic interactions between the cold dark
matter and other matters. So, it provides a strong motivation for us
to consider the collision of two uncharged particles in the
background of a KN black hole. With the motivation, we find the CM
energy can still be unlimited for a pair of uncharged particles
falling freely from rest at infinity and colliding at the horizon of
an extremal black hole with some fine-tunings. For the near-extremal
black hole, we also give a numerical exploration on the CM energy.
The result implies that there always exists a finite upper bound for
the CM energy, which decreases with the increase of the charge $Q$.
In this paper, we neglect the effects of gravitational waves and the
back-reaction.

The paper is organized as follows. In Sec. \ref{CMenergy1}, we will
give a detailed study on the equations of motion for particles. In
Sec. \ref{CMenergy2}, employing the equations of motion for
particles, we will obtain the CM energy for two colliding particles
falling freely from rest at infinity in the background of a KN black
hole. The results show that the CM energy at the horizon can be
unlimited if one of the colliding particles has the critical angular
momentum and the spin $a$ of the black hole satisfies
$\frac{1}{\sqrt{3}}\leq \frac{a}{M}\leq 1$. It is also shown that,
for a near-extremal black hole there always exists a finite upper
bound of CM energy and the bound decreases with the increase of the
charge $Q$. The final section is devoted to a brief summary. We use
the units $c=G=1$ in this paper.

\section{Motion equations of particles in the background of a Kerr-Newman black hole}
\label{CMenergy1}

In this section, we would like to study the equations of motion for
a particle in the background of a KN black hole. First,
let us give a brief review of the black hole background we dealt with.
The KN black hole is described by the metric with the
Boyer-Lindquist coordinates (where we have set the mass $M$ of the
black hole to 1)
\begin{eqnarray}
 ds^{2}&=&\frac{\rho^{2}}{\Delta}dr^{2}+\rho^{2}d\theta^{2}
    +\frac{\sin^{2}\theta}{\rho^{2}}
    \Big[adt-(r^{2}+a^{2})d\phi\Big]^{2}\nonumber\\
    &&-\frac{\Delta}{\rho^{2}}\Big[dt-a \sin^{2}\theta d\phi\Big]^{2},
  \label{KNblackhole}
\end{eqnarray}
where
\begin{eqnarray}
 \Delta&=&r^{2}-2r+a^{2}+Q^{2},\\
 \rho^{2}&=&r^{2}+a^{2}\cos^{2}\theta.
\end{eqnarray}
$Q$ is the charge of {the} black hole, and $a$ is its angular
momentum per unit rest mass {and} $0\leq a\leq 1$. In the case $Q=0$,
{the} metric (\ref{KNblackhole}) describes a Kerr black hole. And in
the case $a=Q=0$, it describes a Schwarzschild black hole. The
4-dimensional electromagnetic potential reads
\begin{eqnarray}
 A_{a}=-\frac{Qr}{\rho^{2}}\bigg[(dt)_{a}-a\sin^{2}\theta(d\phi)_{a}\bigg].
\end{eqnarray}
The horizons for the KN black hole are given by
\begin{eqnarray}
 r_{\pm}=1\pm\sqrt{1-(a^{2}+Q^{2})}.
\end{eqnarray}
Here, the positive sign denotes the outer horizon and the negative
one denotes the inner one. The existence of the horizons requires
\begin{eqnarray}
 a^{2}+Q^{2}\leq 1,
\end{eqnarray}
where "=" corresponds to the extremal black hole with one degenerate
horizon.

Next, we would like to study the equations of motion for a test
particle with mass $\mu$ and charge $q$ in the background of a KN
black hole. The motion of a particle is described by the Lagrangian
\begin{eqnarray}
 \mathcal{L}=\frac{1}{2}g_{\mu\nu}\dot{x}^{\mu}\dot{x}^{\nu}+q
 A_{\mu}\dot{x}^{\mu},
\end{eqnarray}
where a dot over a symbol denotes ordinary differentiation with
respect to an affine parameter $\lambda$. The affine parameter
$\lambda$ is related to the proper time by $\tau=\mu\lambda$, which
is equivalent to the normalizing condition
\begin{eqnarray}
 g_{\mu\nu}\dot{x}^{\mu}\dot{x}^{\nu}=-\mu^{2}.
\end{eqnarray}
For an uncharged particle, $\mu^{2}=1,0,-1$ are corresponded to
timelike, null or spacelike geodesics, respectively. For a massive particle, we have $\mu^{2}=1$. The momenta is
\begin{eqnarray}
  P_{\mu}=\frac{\partial \mathcal{L}}{\partial \dot{x}^{\mu}}
         =g_{\mu\nu}\dot{x}^{\nu}+q A_{\mu}. \label{momenta}
\end{eqnarray}
Thus the Hamiltonian is given by
\begin{eqnarray}
  H=P_{\mu}\dot{x}^{\mu}-\mathcal{L}
   =\frac{1}{2}g^{\mu\nu}(P_{\mu}-q A_{\mu})(P_{\nu}-q A_{\nu}).
   \label{Hamiltonian}
\end{eqnarray}
With the help of (\ref{Hamiltonian}), the Hamilton-Jacobi equation
can be expressed as
\begin{eqnarray}
   \frac{\partial S}{\partial\lambda}
   =H
   =\frac{1}{2}g^{\mu\nu}(P_{\mu}-q A_{\mu})(P_{\nu}-q A_{\nu})
   \label{HJEQ}
\end{eqnarray}
with $S$ the Jacobi action. To solve the Hamilton-Jacobi equation,
we separate the Jacobi action as
\begin{eqnarray}
 S=-\frac{1}{2}\lambda-Et+l\phi+S_{r}(r)+S_{\theta}(\theta),
 \label{action}
\end{eqnarray}
where the parameter $E$ is the energy of the charged particle, and
$l$ is the angular momentum per unit rest mass of the particle in the
$\phi$ direction as measured by an observer at rest at infinity.
$S_{r}$ and $S_{\theta}$ are, respectively, functions of $r$ and
$\theta$. Inserting (\ref{action}) in (\ref{HJEQ}), we obtain
\begin{eqnarray}
 S_{\theta}^{2}+(a E \sin\theta-l\sin^{-1}\theta)^{2}
 +a^{2}\cos^{2}\theta
 =-\Delta S_{r}^{2}
   +\Delta^{-1}\bigg((a^{2}+r^{2})E-al-qQr\bigg)^{2}-r^{2}.
\end{eqnarray}
From the above form, we can see that the left-hand side is only the
function of $\theta$ and the right-hand side is only the function of
$r$. Thus, both sides must be equal to a constant denoted by
$\mathcal{K}$. So, we have
\begin{eqnarray}
 S_{\theta}^{2}&=&\mathcal{K}-(a E \sin\theta-l\sin^{-1}\theta)^{2}
                  -a^{2}\cos^{2}\theta,\\
 \Delta S_{r}^{2}&=&-\mathcal{K}
            +\Delta^{-1}\bigg((a^{2}+r^{2})E-al-qQr\bigg)^{2}-r^{2}.
\end{eqnarray}
Using the relations $P_{r}=\frac{\partial S}{\partial r}$,
$P_{\theta}=\frac{\partial S}{\partial \theta}$ and the Eq.
(\ref{momenta}), we have \cite{Carter,Johnston}
\begin{eqnarray}
 \frac{d\theta}{d\tau}&=&\sigma_{\theta}\frac{\sqrt{\Theta}}{\rho^{2}},\label{thetaequation}\\
 \frac{dr}{d\tau}&=&\sigma_{r}\frac{\sqrt{R}}{\rho^{2}}\label{requation}
\end{eqnarray}
with
\begin{eqnarray}
 \Theta&=&\mathcal{K}-(l-a E)^{2}-\cos^{2}\theta
         \bigg(a^{2}(1-E^{2})+l^{2}\sin^{-2}\theta\bigg),\\
 R&=&P^{2}-\Delta(r^{2}+\mathcal{K}),\\
 P&=&E(r^{2}+a^{2})-la-qQr.
\end{eqnarray}
The sign functions $\sigma_{r}=\pm$ and $\sigma_{\theta}=\pm$ are
independent from each other. Using the relations
$P_{t}=\frac{\partial S}{\partial t}$, $P_{\phi}=\frac{\partial
S}{\partial \phi}$ and Eq. (\ref{momenta}), we get
\begin{eqnarray}
 -E&=&g_{tt}\dot{t}+g_{t\phi}\dot{\phi}+q A_{t},\\
 l&=&g_{\phi t}\dot{t}+g_{\phi\phi}\dot{\phi}+q A_{\phi}.
\end{eqnarray}
Solving these equations, we get \cite{Carter,Johnston}
\begin{eqnarray}
 \frac{dt}{d\tau}&=&-\frac{a}{r^{2}}(a E\sin^{2}\theta-l)
                  +\frac{(r^{2}+a^{2})}{\rho^{2}\Delta}P,\label{tequation}\\
 \frac{d\phi}{d\tau}&=&-\frac{(aE\sin^{2}\theta-l)}{\rho^{2}\sin^{2}\theta}
     +\frac{a}{\rho^{2}\Delta}P. \label{phiequation}
\end{eqnarray}
Here, we have obtain equations of motion for a particle. On the
equatorial plane $(\theta=\frac{\pi}{2})$, the equations are reduced
to
\begin{equation}
  \left.
   \begin{aligned}[c]
 &\frac{dt}{d\tau}=\frac{-a(a E-l)\Delta+(r^{2}+a^{2})P}{r^{2}\Delta},\label{4S}\\
 &\frac{dr}{d\tau}=-\frac{\sqrt{R}}{r^{2}},\\
 &\frac{d\theta}{d\tau}=0,\\
 &\frac{d\phi}{d\tau}=\frac{(l-aE)\Delta+aP}{r^{2}\Delta},
    \end{aligned}
  \right.
\end{equation}
where we take $\sigma_{r}=-1$. Note that the motion of a particle on
the equatorial plane in the KN metric is
completely determined by Eq. (\ref{4S}).

\section{Center-of-mass energy for a Kerr-Newman black hole}
\label{CMenergy2}

In this section, we will study the CM energy of the collision for
two particles moving on the equatorial plane of a KN black hole. Let
us now consider that two charged particles with the same rest mass
$m_0$ are at rest at infinity ($E=m_{0}$), then they approach the
black hole and collide at some radius $r$. We assume that the two
particles have angular momenta and charges ($l_{1}$, $q_{1}$) and
($l_{2}$, $q_{2}$), respectively. Taking into account that the
background is curved, the energy in the center-of-mass frame for
this collision should be computed with \cite{Banados}
\begin{eqnarray}
 E_{\text{cm}}=\sqrt{2}m_{0}\sqrt{1-g_{\mu\nu}U_{(1)}^{\mu}U_{(2)}^{\nu}},
  \label{energy}
\end{eqnarray}
where $U_{(1)}^{\mu}$ and $U_{(2)}^{\nu}$ are the 4-velocities of
the two particles, which can be straightforwardly calculated from
(\ref{4S}) and are
\begin{eqnarray}
 U_{(1)}^{\mu}=\bigg(\frac{a(l_{1}-a)\Delta+(r^{2}+a^{2})P(q_{1},l_{1})}{r^{2}\Delta},
               -\frac{\sqrt{R}}{r^{2}},0,
               \frac{(l_{1}-a)\Delta+aP(q_{1},l_{1})}{r^{2}\Delta}\bigg),\\
 U_{(2)}^{\mu}=\bigg(\frac{a(l_{2}-a)\Delta+(r^{2}+a^{2})P(q_{2},l_{2})}{r^{2}\Delta},
               -\frac{\sqrt{R}}{r^{2}},0,
               \frac{(l_{2}-a)\Delta+aP(q_{2},l_{2})}{r^{2}\Delta}\bigg).
\end{eqnarray}
Here, we take $E=1$ for simplicity. With the help {of}
(\ref{energy}), we obtain the CM energy for the collision:
\begin{eqnarray}
 \bigg(\frac{E_{\text{cm}}}{\sqrt{2}m_{0}}\bigg)^{2}=-\frac{H}{r^{2}\Delta}
 \label{KNenergy}
\end{eqnarray}
with $H$ given by
\begin{eqnarray}
 H&=&-2r^{4}+r^{3}\big[2+Q(q_{1}+q_{2})\big] \nonumber\\
    &&-r^{2}(2a^{2}+Q^{2}-l_{1}l_{2}+Q^{2}q_{1}q_{2})-2a^{2}r\nonumber\\
 &&+2r\big[a(l_{1}+l_{2})-l_{1}l_{2}\big] +Q^{2}(a-l_{1})(a-l_{2})\nonumber\\
 &&    +aQr\big[a(q_{1}+q_{2})-(l_{2}q_{1}+l_{1}q_{2})\big]\nonumber\\
   &&+\sqrt{(a^{2}+r^{2}-al_{1}-Qrq_{1})^{2}-\Delta(r^{2}+(a-l_{1})^{2})}\nonumber\\
   &&  \times \sqrt{(a^{2}+r^{2}-al_{2}-Qrq_{2})^{2}-\Delta(r^{2}+(a-l_{2})^{2})}.\nonumber\\
\end{eqnarray}
Note that (\ref{KNenergy}) is invariant under the interchange $
l_{1}\leftrightarrow l_{2}$ and $q_{1}\leftrightarrow q_{2}$. On the
other hand, in the case $Q=q_{1}=q_{2}=0$, the CM energy
(\ref{KNenergy}) for two charged particles in the background of a KN
black hole will reduce to the one for two uncharged particles in the
background of a Kerr black hole given in \cite{Banados}, as it is
expected. In fact, we need only the condition $Q=0$ to obtain the CM
energy in the background of a Kerr black hole, which indicates that
the charge of the collision particles has no influence on the CM
energy in the background of an uncharged black hole. We keep in mind
that black holes are surrounded by relic cold dark matter density
spikes. The CM energy of massive cold dark matter particles
colliding near the black hole may reach a high CM energy, which
could provide a probe to the high energy physics. It is also thought
that the cold dark matter has no electromagnetic interactions with
other matters. So, we consider that two uncharged cold dark matter
particles collide in the background of a KN black hole. Thus, we
take $q_{1}=q_{2}=0$. Then the CM energy can be read from
(\ref{KNenergy}):
\begin{eqnarray}
 \bigg(\frac{E_{\text{cm}}}{\sqrt{2}m_{0}}\bigg)^{2}=-\frac{K}{r^{2}\Delta},
 \label{KN2blackhole}
\end{eqnarray}
where $K$ is
\begin{eqnarray}
 K&=&-2r^{4}+2r^{3}-r^{2}(2a^{2}+Q^{2}-l_{1}l_{2})-2a^{2}r\nonumber\\
   &&
       +2r\big[a(l_{1}+l_{2})-l_{1}l_{2}\big]+Q^{2}(a-l_{2})(a-l_{2}) \nonumber\\
   &&+\sqrt{(a^{2}+r^{2}-al_{1})^{2}-\Delta[r^{2}+(a-l_{1})^{2}]}\nonumber\\
   &&
       \times \sqrt{(a^{2}+r^{2}-al_{2})^{2}-\Delta[r^{2}+(a-l_{2})^{2}]}.
\end{eqnarray}
Clearly, the result confirms that the charge $Q$ of the black hole
indeed has influence on the CM energy.

\begin{figure*}
\centerline{\subfigure[]{\label{Veff}
\includegraphics[width=8cm,height=6cm]{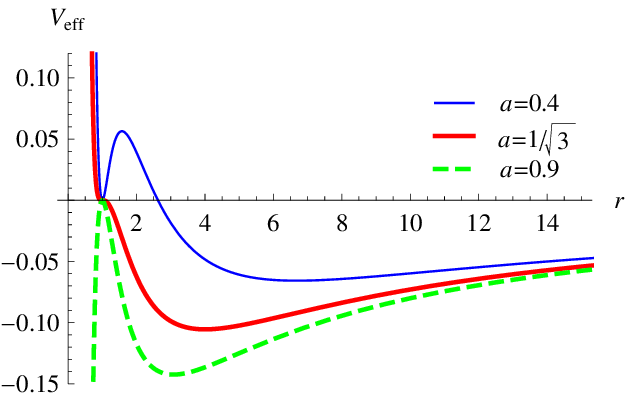}}
\subfigure[]{\label{ECM}
\includegraphics[width=8cm,height=6cm]{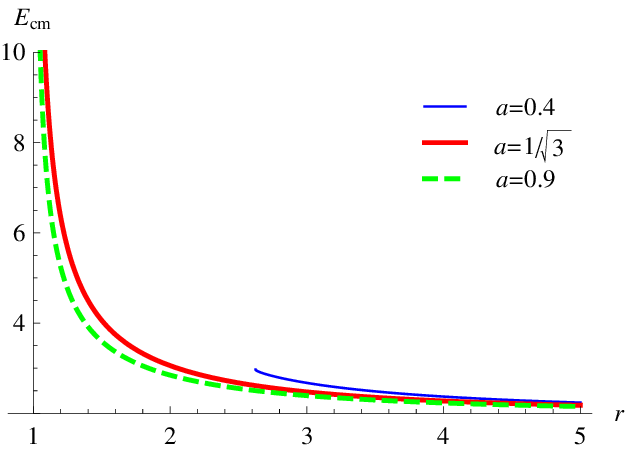}}}
\caption{For an extremal KN black hole (a) shows the effective
potential $V_{\text{eff}}$ vs $r$ with angular momentum
$l=l_{\text{c}}=\frac{1+a^{2}}{a}$. (b) shows the behavior of the CM
energy $E_{\text{cm}}$ with $m_{0}=1$ and $l_{1}=l_{\text{c}}$,
$l_2=-2$.}
\end{figure*}

Next, it is worthwhile to study the properties of (\ref{KN2blackhole}) as
the radius $r$ approaches to the horizon $r_+$ of an extremal black hole. Since the black hole
considered here is an extremal one, we have $Q=\sqrt{1-a^{2}}$. Note
that the horizon is always at $r_{+}=1$ for any spin $a$. It is
clear that the denominator of $E_{\text{cm}}$ in
(\ref{KN2blackhole}) vanishes at $r=r_{+}$. Then it is naive to
obtain the result that $E_{\text{cm}}$ diverges at the horizon. In
fact, the numerator also vanishes at that point. The limiting value
of $E_{\text{cm}}$ at $r=r_+$ can be calculated as follows:
\begin{eqnarray}
 E_{\text{cm}}(r\rightarrow r_{+})=2 m_{0}
     \sqrt{1+\frac{(l_{1}-l_{2})^2}{(l_{1}-l_{\text{c}})
   (l_{2}-l_{\text{c}})}\frac{l_{\text{c}}}{4a}}, \nonumber \\
   \;\;\;(a^{2}+Q^{2}=1).
                    \label{LKN}
\end{eqnarray}
Clearly, the value of $E_{\text{cm}}$ is indeed finite for generic
values of $l_{1}$ and $l_{2}$. However, when $l_{1}$ or $l_{2}$
takes the critical angular momentum
\begin{eqnarray}
 l_{\text{c}}=\frac{1+a^{2}}{a}, \label{lc}
\end{eqnarray}
the CM energy $E_{\text{cm}}$ will be unlimited, which means that
the particles can collide with arbitrarily high CM energy at the
horizon. Thus, the result may provide an effective way to probe the
Planck-scale physics in the background of an extremal KN black hole.
Compared with the result for the Kerr black hole \cite{Banados}, the
spin $a$ of the black hole here can deviate from its maximal value
to obtain an arbitrarily high CM energy. However, we need to make
sure that the particle with critical angular momentum $l_{\text{c}}$
can reach the horizon.

As we mentioned before, to obtain an arbitrarily high CM energy, one
of the colliding particles should have critical angular momentum
$l_{\text{c}}$. Here we first examine the critical angular momentum
(\ref{lc}). When the spin $a=1$, we get $l_{\text{c}}=2$, which is
just the critical angular momentum in the case of an extremal Kerr
black hole \cite{Banados}. However, for the case $a=0$,
$l_{\text{c}}$ is divergent and the CM energy is
\begin{eqnarray}
 E_{\text{cm}}(r\rightarrow r_{+})=2 m_{0}
     \sqrt{1+\frac{(l_{1}-l_{2})^2}{4}},
     \label{LRN}
\end{eqnarray}
which implies that, in order to get a very high CM energy, one of
the colliding particles should have very large angular momentum.
However, a particle with very large angular momentum can not reach
the horizon if it falls freely from rest at infinity. So there must
exist a range for the spin $a$ to ensure that the particle with critical
angular momentum $l_{\text{c}}$ reaches the horizon of the black hole.
Next, we will determine the range of the spin $a$ by the effective
potential method. The effective potential for a particle with
critical angular momentum $l_{\text{c}}$ on the equatorial plane of
an extremal black hole is
\begin{eqnarray}
 V_{\text{eff}}=-\frac{1}{2}\bigg(\frac{dr}{d\tau}\bigg)^{2} 
        =-\frac{(r-1)^2 \left(r-r_{\text{c}}\right)}{r^4}
\end{eqnarray}
with $r_{\text{c}}=\frac{1-a^2}{2 a^2}$. As expected, the effective
potential $V_{\text{eff}}$ approaches 0 at infinity. Here, we can
get a condition for the particle falling freely from rest at
infinity to reach the horizon:
\begin{eqnarray}
 V_{\text{eff}} \leq 0 \quad \mbox{for any}\quad  r\geq 1,
\end{eqnarray}
which is equivalent to
\begin{eqnarray}
 r_{\text{c}}\leq 1. \label{rc}
\end{eqnarray}
Solving Eq. (\ref{rc}), we get the range for the spin $a$ of the
black holes:
\begin{eqnarray}
 \frac{1}{\sqrt{3}}\leq a\leq 1,
\end{eqnarray}
which means that for an extremal black hole with the spin
$a\in(\frac{1}{\sqrt{3}},\;1)$, the particle with critical angular
momentum $l_{\text{c}}$ can reach the horizon of the black hole. We
can also determine the range of the angular momentum for a fixed
spin $a$ with the same method. However, it can not be written in a
closed form for an arbitrary spin $a$. Here, we give some results:
the range of the angular momentum is $(-3.9539,\;2.6471)$ for
$a=0.4$, $(-4.2185,\;2.3094)$ for $a=\frac{1}{\sqrt{3}}$ and
$(-4.6864,\;2.0111)$ for $a=0.9$. So, for a fixed spin
$a\in(\frac{1}{\sqrt{3}},\;1)$, if $l_{1}=l_{\text{c}}$ and $l_{2}$
is in a proper range, the CM energy will be unlimited. We plot the
effective potential $V_{\text{eff}}$ in Fig. \ref{Veff} for the spin
$a=0.4, \frac{1}{\sqrt{3}}$ and 0.9, respectively. Clearly, for
$a=0.4$, the effective potential $V_{\text{eff}}$ is positive near
the horizon $r_{+}=1$, so the particle can not reach the horizon in
this case. For $a=\frac{1}{\sqrt{3}}$ and 0.9, the effect potential
$V_{\text{eff}}$ is negative when $r>r_{+}=1$. So, the particle can
reach the horizon in both cases. We also plot the CM energy
$E_{\text{cm}}$ of collision in Fig. \ref{ECM} for
$l_{1}=l_{\text{c}}$ and $l_{2}=-2$. For the case
$a=0.4<\frac{1}{\sqrt{3}}$, the CM energy only exists for $r>2.625$.
This is because the collision for the two colliding particles with
angular momenta $l_{1}=l_{\text{c}}$ and $l_{2}=-2$ can not take
place at $r<2.625$. For the spin $a=\frac{1}{\sqrt{3}}$ and 0.9, the
CM energy is divergent at the horizon $r_{+}=1$.

As noted above, we show that an arbitrarily high CM energy can be
obtained when the collision takes place at the horizon of an extremal
KN black hole with $l=l_{\text{c}}$ and $a\in (1/\sqrt{3},\;1)$.
This scenario is an idealized one, because the proper time for a
particle with the critical angular momentum $l_{\text{c}}$ to
approach the horizon of an extremal black hole from infinity is
infinite. Thus this collision process does not take place in the real
world. However, for the case of a near-extremal black hole, the
proper time for the particle to reach the horizon is a finite value
even though it is very large. So, it seems worthwhile to consider a
near-extremal black hole. The CM energy at the outer horizon $r_{+}$
for a near-extremal black hole is found to be
\begin{eqnarray}
 E_{\text{cm}}(r\rightarrow r_{+})=2 m_{0}
     \sqrt{1+\frac{(l_{1}-l_{2})^2}{(l_{1}-l'_{\text{c}})
   (l_{2}-l'_{\text{c}})}\frac{l'_{\text{c}}}{4a}}, \label{outCM}
\end{eqnarray}
where
\begin{eqnarray}
 l'_{\text{c}}=\frac{2+2\sqrt{1-a^{2}-Q^{2}}-Q^{2}}{a}. \label{lc2}
\end{eqnarray}
The form is the same as (\ref{LKN}) with the replacement
$l_{\text{c}}\rightarrow l'_{\text{c}}$. Here, we denote the small
parameter $\epsilon=a_{\text{max}}-a$ with
$a_{\text{max}}=\sqrt{1-Q^{2}}$. For fixed charge $Q$ and
$\epsilon$, the range $(l_{\text{min}},\;l_{\text{max}})$ of angular
momentum for the particles to reach the horizon can be determined
numerically with the effective potential $V_{\text{eff}}$ for a near-extremal black hole. For a angular momentum
$l\in(l_{\text{min}},\;l_{\text{max}})$, we can get a negative
$V_{\text{eff}}(l)$ for $r>r_{+}$. However, for arbitrary charge $Q$
and $\epsilon$, we find that, within a small range near the horizon
$r_{+}$, the effective potential $V_{\text{eff}}(l'_{\text{c}})$ is always positive,
which means the angular momentum $l'_{\text{c}}$ does not
lie in the range $(l_{\text{min}},\;l_{\text{max}})$. So the CM
energy $E_{\text{cm}}$ in (\ref{outCM}) is not divergent. Thus the
CM energy is finite for arbitrary charge $Q$ and spin $a$.
Considering that one of the colliding particles has the maximum
angular momentum $l_{\text{max}}$ and another one has the minimum
angular momentum $l_{\text{min}}$, we obtain the CM energy per unit rest
mass for different $Q$ and $\epsilon$. The result is shown in Table
\ref{TableCMenergy}. From it, we can see that, for a KN black hole
with spin $a$ less than $a_{\text{max}}$ there will be an upper
bound for the CM energy. It is also suggested that the CM energy
grows very slowly as the maximally spinning case
($\epsilon\rightarrow 0$) is approached. For fixed parameter
$\epsilon$, the value of CM energy decreases with the increase of
the charge $Q$. For the case $Q=0$, it describes a Kerr black hole
and the result shown in Table \ref{TableCMenergy} is exactly
consistent with \cite{Jacobson}.

\begin{table}[h]
\begin{center}
\caption{The CM energy per unit rest mass $\frac{E_{\text{cm}}}{m_{0}}$
for a KN black hole with spin $a=a_{\text{max}}-\epsilon$ and
$l_{1}=l_{\text{max}}$, $l_{2}=l_{\text{min}}$.}\label{TableCMenergy}
\begin{tabular}{c c c c c c }
  \hline
  \hline  
    & $\epsilon$=0.1 & $\epsilon$=0.05 & $\epsilon$=0.01 & $\epsilon$=0.001 & $\epsilon$=0.0001 \\
  \hline $Q$=0\;\; &   6.901 & 8.244 & 12.54 & 22.63 & 40.49 \\
 $Q$=0.1 & 6.894 & 8.234 & 12.51 & 22.59 & 40.40 \\
 $Q$=0.2 & 6.875 & 8.203 & 12.45 & 22.44 & 40.12 \\
 $Q$=0.3 & 6.842 & 8.150 & 12.33 & 22.19 & 39.64 \\
 $Q$=0.4 & 6.794 & 8.073 & 12.16 & 21.82 & 38.93 \\
 $Q$=0.5 & 6.730 & 7.967 & 11.93 & 21.30 & 37.94 \\
 $Q$=0.6 & 6.647 & 7.826 & 11.60 & 20.57 & 36.54 \\
 $Q$=0.7 & 6.539 & 7.636 & 11.14 & 19.49 & 34.44 \\
$Q$=0.8 & 6.398 & 7.367 & 10.42 & 17.57 & 30.35 \\
  \hline
\end{tabular}
\end{center}
\end{table}

In order to obtain a Planck-scale CM energy ($E_{\text{Pl}} \sim 10^{19}$ Gev), we would like to study how much the tolerance on the critical angular momentum $l_{\text{c}}$ and on the black hole parameter. First, we consider the case that the black hole is still an extremal one, but there exists a small tolerance $\delta l$ on the
critical angular momentum $l_{\text{c}}$. For simplicity, we choose
$l_{1}=l_{\text{c}}-\delta l$ and $l_{2}=0$. The rest mass for the
colliding particle is considered to $m_{0}\sim 1$ GeV, just like the
mass of a neutron. Then with the help of (\ref{LKN}), we get a
approximate $\delta l$:
\begin{eqnarray}
 \delta l \approx
            \frac{l_{\text{c}}^{2}}{a}\left(\frac{m_{0}}{E_{\text{Pl}}}\right)^{2} 
            \sim 10^{-37}.
\end{eqnarray}
Note that we have considered that $a\in (\frac{1}{\sqrt{3}},1)$.
Now, we would like to estimate the tolerance on the extremal black
hole parameters to achieve the Planck-scale energy. Here we consider
that one of the colliding particles has the critical angular momentum $l_{\text{c}}$,
but the black hole is a near-extremal one. Here, we denote the
tolerance $\epsilon=a_{\text{max}}-a \ll 1$, and suppose that
$l_{1}=l_{\text{c}}$ and $l_{2}=0$. Then, for $m_{0}\sim 1$ GeV,
with the CM energy (\ref{outCM}), we have
\begin{eqnarray}
 \epsilon \approx
            \frac{l_{\text{c}}^{4}}{8a}\left(\frac{m_{0}}{E_{\text{Pl}}}\right)^{4} 
             \sim 10^{-76}. \label{EEP}
\end{eqnarray}
Here, we have shown that to achieve the Planck-scale energy, if the
black hole is an extremal one, then the tolerance on the critical
angular momentum is $\delta l\sim 10^{-37}$ and if one of the
colliding particles has the critical angular momentum $l_{\text{c}}$
but the black hole is a near-extremal one, then the tolerance on the
black hole parameter is $\epsilon\sim 10^{-76}$. Replacing
Planck-scale energy $E_{\text{Pl}}$ with an arbitrary energy
$E_{\text{cm}}$, (\ref{EEP}) can be reexpressed as
\begin{eqnarray}
 \left(\frac{E_{\text{cm}}}{m_{0}}\right)\sim
 \frac{l_{\text{c}}}{\sqrt[4]{8a}}\epsilon^{-1/4}. \label{ourE}
\end{eqnarray}
Comparing with (\ref{ECMPP}), we find they have the same order
$\epsilon^{-1/4}$. However, our formula (\ref{ourE}) is only an
approximation, the more exact result can be found in Table
\ref{TableCMenergy} for different charge $Q$ and $\epsilon$.

\section{Summary}

In this paper, we have investigated the collision of two uncharged
particles (which could be thought to be the cold dark matter
particles) falling freely from rest at infinity in the background of
a KN black hole. It is pointed out by BSW \cite{Banados} that the CM
energy of collision for two particles in the background of an
extremal Kerr black hole can approach to an arbitrarily high value if
one of the particles has angular momentum $l=2$. Our results show
that when extended to the KN black hole case, an unlimited CM energy
requires three conditions: (1) the collision takes place at the
horizon of an extremal black hole; (2) one of the colliding particles
has critical angular momentum $l=\frac{1+a^2}{a}$; (3) the spin $a$
of the extremal black hole satisfies $\frac{1}{\sqrt{3}}\leq a\leq
1$. Compared with the Kerr black hole, to obtain an arbitrarily high
CM energy, besides the conditions that the black hole is an extremal
black hole and one of the colliding particles has critical angular
momentum, there still exists a restriction on the value of the spin
$a$ of the KN black hole, which is a significant difference between
the two black holes. For a near-extremal black hole, we also find
that there always exists an upper bound for the CM energy, which
decreases with the increase of the charge $Q$. For an extremal black
hole, in order to obtain the Planck-scale energy, the tolerance on
the critical angular momentum should be $\delta l\sim 10^{-37}$. On
the other hand, if one of the colliding particles has the critical
angular momentum $l_{\text{c}}$, then the tolerance on the black
hole parameter is $\epsilon\sim 10^{-76}$. However, if the particle
does not fall freely from rest at infinity
\cite{Grib,Pavlov,Harada}, the unlimited CM energy may be
approached. In future work, we will explore the CM energy for the
collision of charged particles taking place in a nonextremal black
hole background.

\section*{Acknowledgement}

This work was supported by the Program for New Century Excellent
Talents in University, the Huo Ying-Dong Education Foundation of
Chinese Ministry of Education (No. 121106), the National Natural
Science Foundation of China (No. 11075065), and the Fundamental Research Funds for the
Central Universities (No. lzujbky-2009-54 and No. lzujbky-2009-163).

\end{document}